\documentclass[useAMS,usenatbib]{mn2e}
\usepackage{graphicx}
\title[Long-term evolution of dim isolated neutron stars]{Long-term evolution of dim isolated neutron stars}
\author[\"{U}. Ertan, \c{S}. \c{C}al{\i}\c{s}kan, O. Benli \& M.A. Alpar ]{\"{U}. Ertan\thanks{E-mail:unal@sabanciuniv.edu}, \c{S}. \c{C}al{\i}\c{s}kan, O. Benli \& M.A. Alpar\\
Sabanc\i\ University, 34956, Orhanl\i\, Tuzla, \.Istanbul, Turkey}
\begin{document}

\date{4 August 2014 }


\maketitle

\newcommand {\gtrsim} {\ {\raise-.5ex\hbox{$\buildrel>\over\sim$}}\ }
\newcommand {\lesssim} {\ {\raise-.5ex\hbox{$\buildrel<\over\sim$}}\ }

\def\la{\raise.5ex\hbox{$<$}\kern-.8em\lower 1mm\hbox{$\sim$}}
\def\ga{\raise.5ex\hbox{$>$}\kern-.8em\lower 1mm\hbox{$\sim$}}
\def\be{\begin{equation}}
\def\ee{\end{equation}}
\def\ba{\begin{eqnarray}}
\def\ea{\end{eqnarray}}
\def\m{\mathrm}
\def\d{\partial}
\def\R{\right}
\def\L{\left}
\def\a{\alpha}
\def\acold{\alpha_\mathrm{cold}}
\def\ahot{\alpha_\mathrm{hot}}
\def\Mdotstar{\dot{M}_\ast}
\def\Omegastar{\Omega_\ast}
\def\OmegaK{\Omega_{\mathrm{K}}}
\def\Mdotin{\dot{M}_{\mathrm{in}}}
\def\Mdot{\dot{M}}
\def\Edot{\dot{E}}
\def\Pdot{\dot{P}}
\def\Msun{M_{\odot}}
\def\Lin{L_{\mathrm{in}}}
\def\Lcool{L_{\mathrm{cool}}}
\def\Rin{R_{\mathrm{in}}}
\def\rin{r_{\mathrm{in}}}
\def\rlc{r_{\mathrm{LC}}}
\def\rout{r_{\mathrm{out}}}
\def\rco{r_{\mathrm{co}}}
\def\Rout{R_{\mathrm{out}}}
\def\Ldisk{L_{\mathrm{disk}}}
\def\Lx{L_{\mathrm{x}}}
\def\Md{M_{\mathrm{d}}}
\def\NH{N_{\mathrm{H}}}
\def\cs{c_{\mathrm{s}}}
\def\dEb{\delta E_{\mathrm{burst}}}
\def\dEx{\delta E_{\mathrm{x}}}
\def\Bstar{B_\ast}
\def\Bb{\beta_{\mathrm{b}}}
\def\Be{\beta_{\mathrm{e}}}
\def\Bp{B_{\mathrm{p}}}
\def\Rc{\R_{\mathrm{c}}}
\def\rA{r_{\mathrm{A}}}
\def\rp{r_{\mathrm{p}}}
\def\Tp{T_{\mathrm{p}}}
\def\dMin{\delta M_{\mathrm{in}}}
\def\dM*{\delta M_*}
\def\Teff{T_{\mathrm{eff}}}
\def\Tirr{T_{\mathrm{irr}}}
\def\Firr{F_{\mathrm{irr}}}
\def\Tcrit{T_{\mathrm{crit}}}
\def\P0min{P_{0,{\mathrm{min}}}}
\def\Av{A_{\mathrm{V}}}
\def\ah{\alpha_{\mathrm{hot}}}
\def\ac{\alpha_{\mathrm{cold}}}
\def\tc{\tau_{\mathrm{c}}}
\def\p{\propto}
\def\m{\mathrm}
\def\fast{\omega_{\ast}}
\def\Alfven{Alfv$\acute{e}$n}
\def\418{SGR 0418+5729}
\def\142{AXP 0142+61}
\def\Caliskan{\c{C}al{\i}\c{s}kan~}
\def\qleft{\textquotedblleft}
\def\qright{\textquotedblright~}
\label{firstpage}

\begin{abstract}
The dim isolated neutron stars (XDINs) have periods in the same range as the anomalous X-ray pulsars (AXPs) and the soft gamma-ray repeaters (SGRs). We apply the fallback disk model, which explains the period clustering and other properties of AXP/SGRs, to the six XDINs with measured periods and period derivatives. Present properties of XDINs are obtained in evolutionary scenarios with surface dipole magnetic fields $B_0 \sim 10^{12}$ G. The XDINs have gone through an accretion epoch with rapid spin-down earlier, and have emerged in their current state, with the X-ray luminosity provided by neutron star cooling and no longer by accretion. Our results indicate that the known XDINs are not likely to be active radio pulsars, as the low $B_0$, together with their long periods place these sources clearly below the \textquotedblleft death valley\textquotedblright. 
\end{abstract}


\section{INTRODUCTION}

What are the physical conditions leading to formation of different populations of young, isolated (single) neutron stars, namely anomalous X-ray pulsars (AXPs), soft gamma-ray repeaters (SGRs), dim isolated neutron stars (XDINs), central compact objects (CCOs) and rotating radio transients (RRATs)? All these systems involve a single, young or middle-aged neutron star (see Mereghetti 2011a for a recent review of isolated neutron stars). Despite striking differences, like energetic soft gamma bursts peculiar to SGRs and AXPs, there are also striking similarities, like the periods of AXP/SGRs and XDINs being clustered in the same range ($2 - 12$ s). Considering the estimated birth rates of these objects together with the galactic supernova rate, it is likely that there are evolutionary connections between some of these populations (Keane \& Kramer 2008). Recent efforts concentrate on the unification of the long-term X-ray luminosity and the rotational evolution of these neutron star systems in a single picture. With the assumption that these sources are evolving in vacuum with dipole torques, their surface dipole fields, B, inferred from the dipole torque formula range from $\sim 10^{10}$ G for CCOs to more than $10^{15}$ G for AXP/SGRs. In this picture, the diversity of evolutionary paths is attributed to the differences in the initial dipole and the crustal toroidal fields of the sources (Kaspi 2010). Sources with magnetar dipole fields ($B > 10^{14}$ G) are posited to go through a rapid field decay, which is required for the model to explain the recently discovered  so-called \qleft low-B magnetars\qright (Turolla et al. 2011, Rea et al. 2012). Evolution of the X-ray luminosities of these sources, which are much higher than the rotational powers, are suggested to be governed by the field decay history of the neutron stars depending on the initial crustal and dipole field strengths (Vigano et al. 2013). In this model, the apparently missing link between CCOs and other classes was suggested to be due to field burial to the crust by the accretion of the supernova matter in the early phase of evolution (Vigano \& Pons 2012). The timescale for the subsequent reemergence and the growth of the field to its original strength is estimated to be $10^3 - 10^5$ yr depending on the initial conditions by the same authors. The dipole field's decay is supposed to proceed after the initial burial and re-emergence. If there are fallback disks around these systems the picture is rather different. It was proposed by Alpar (2001) that the source properties, in particular the period clustering, could be explained by the presence (or absence) and the properties of fallback disks around these systems. Fallback disk models do not involve burial, re-emergence or decay of the dipole field, which is taken to be constant at its initial value.
 
Neutron stars with  fallback disks and  conventional dipole magnetic fields $(B= 10^{12} - 10^{13}$ G) could evolve into the  X-ray luminosity, period and period derivative range of AXP/SGRs on timescales of $10^3 -$ a few $10^5$ yr (Ertan et al. 2009). The infrared (IR) and optical emission properties of these sources in the quiescent state are consistent with the emission from an irradiated active disk (Ertan et al. 2007, Ertan \& \Caliskan 2006). The model fits to optical and IR data also constrain the dipole field strength to below $\sim 10^{13}$ G on the surface of the star (Ertan et al. 2007). Recently discovered  \textquotedblleft low-B magnetars\textquotedblright~SGR 0418+5729 (Rea et al. 2013) and Swift J1822.3$-$1606 (Scholz et al. 2012) clearly showed that the SGR bursts do not require magnetar strength dipole fields. This suggests that the SGR bursts could be powered by the quadrupole or higher multipole fields which are localized close to the surface and do not affect the rotation history of the star. The inner disk applies torque on the star through interaction with the large-scale dipole component of the magnetic field. The X-ray luminosity and the rotational properties of the \qleft low-B magnetars\qright and the so-called \textquotedblleft high-B radio pulsar\textquotedblright~PSR J1734$-$3333, including its anomalous braking index n = 0.9 $\pm$ 0.2 (Espinoza et al. 2011), can also be reproduced by the same model and with $B = 1 - 2 \times 10^{12}$ G. As an independent and complementary support for this model, recent analyses show that the high energy spectra of AXP/SGRs could be produced in their accretion columns consistently with the accretion rates of these sources inferred from the X-ray luminosities (Tr\"umper et al. 2010, 2013; Kylafis, Tr{\"u}mper \& Ertan 2014).       

These results obtained from the application of the same model with similar basic parameters to sources with rather different observational properties encourages us to investigate the evolution and possible evolutionary links between AXP/SGRs (see Mereghetti 2008; 2011b for recent reviews) and other young neutron star populations. This will help us understand the differences in the initial conditions that lead to emergence of different classes of neutron stars. In particular, understanding whether there is a relation between the disk masses and the dipole field strengths requires detailed investigations of AXP/SGRs and other young neutron star populations.  In the present work, we concentrate on the dim isolated neutron stars (XDINs). 

At present, there are seven known  dim isolated neutron stars (Haberl 2007; Turolla 2009). All these sources lie within a distance of $\sim 400$ pc. Since the solar neighborhood is part of the Gould Belt, a ring of young stellar systems, the relatively high rate of supernova events in the Gould Belt should be taken into account in estimating the birth rate of XDINs. In the vicinity of the Sun, about two thirds of the neutron stars are born in the Gould Belt, while the remaining fraction belongs to the Galactic disk (Grenier 2000). Assuming the ages of XDINs are $ \sim 1$ Myr,  simple statistical calculations give a galactic birth rate of $\sim 1$ century$^{-1}$ (Popov, Turolla \& Possenti 2006). The thermal X-ray luminosities of XDINs are in the $10^{31} - 10^{32}$ erg s$^{-1}$ range. The ages corresponding to these luminosities on the theoretical cooling curves are a few $10^5$ yr. Kinematic ages of four XDINs estimated from the space velocities and likely birth places are, with large uncertainties, in the range of  $\sim 0.1 - 1$ Myr (Motch et al. 2009; Tetzlaff et al. 2010; Mignani et al. 2013). These kinematic ages and the cooling ages estimated from the X-ray luminosities of XDINs, imply that the characteristic ages, estimated assuming isolated pulsar spin-down, $P / 2 \Pdot\sim 1 - 4$ Myr, could be significantly greater than the true ages of these sources.    

XDINs do not show pulsed radio emission. Weak radio emission from two sources were reported (Malofeev, Malov \& Teplykh 2007), but not confirmed yet (Turolla 2009; Haberl 2007). Strength of the surface dipole fields of XDINs inferred from the dipole torque formula are $\sim 10^{13} - 10^{14}$ G. If these sources are indeed evolving in vacuum with dipole torques, many of them should be active as radio pulsars. Nondetection of radio pulses from these sources might be due to narrow beaming angles of long-period system. Nevertheless it is not clear why a large population of galactic XDINs with rotational properties of the known sources do not show up as radio pulsars. Alternatively, these sources may not have sufficiently strong dipole fields for radio emission. Our results support the latter possibility. In the present work, we try to explain the long-term evolution of XDINs in the frame of the fallback disk model as applied earlier to AXP/SGRs and PSR J1734$-$3333. We also discuss the radio properties of XDINs based on the predictions of our evolutionary model. We briefly describe the model in Section 2. The results of the simulations are given in Section 3. We summarize our conclusions in Section 4. 

\section{MODEL}

We use the code developed to investigate the long-term evolution of AXPs and SGRs (see Ertan \& Erkut 2008; Ertan et al. 2009; Alpar et al. 2011; \Caliskan et al. 2013 for details and applications). We examine the period, the period derivative and the total X-ray luminosity evolution of the model sources, tracing the initial conditions, namely the initial period, $P_0$,   strength of the magnetic dipole field on the pole of the star, $B_0$, and the initial disk mass, $\Md$. In addition to these initial parameters, the evolution is also affected by the irradiation efficiency $C$ and the minimum critical temperature, $\Tp$, for the disk to be viscously active. The magneto-rotational instability (Balbus \& Hawley 1991) which generates the turbulent viscosity needed for the disk to transport mass and angular momentum will not work at temperatures below $\Tp$, because the ionization fraction becomes too small. 

The \Alfven ~radius of the disk could be written as 
\be
\rA \simeq (G M)^{-1/7}~\mu^{4/7}~ \Mdotin^{-2/7}         
\ee
(Lamb, Pethick \& Pines 1973, Davidson \& Ostriker 1973) where $G$ is the gravitational constant, $M$ and $\mu$ are the mass and magnetic dipole moment of the neutron star, and  $\Mdotin$ is the mass-flow rate arriving at the inner disk radius, $\rin$. When $\rA$ is less than the light cylinder radius, $\rlc = c / \Omegastar$, we take $\rin = \rA$. Accretion will take place in this regime. Typically for the sources we consider, in the accretion phase  $\rin < \rlc$ and the inner disk radius is greater than the co-rotation radius, $\rin > \rco = (G M / \Omega^2)^{1/3}$. The star is in the propeller phase, spinning down under the disk torques while accreting. Over the long-term evolution, $\rlc$ increases with decreasing angular frequency of the neutron star, $\Omegastar$, while $\rA$ increases with decreasing $\Mdotin$.  When $\rA$ calculated by Equation (1) is found to be greater than the current value of $\rlc$, we set  $\rin = \rlc$. In this \textquotedblleft tracking phase\textquotedblright, we assume that there is no accretion onto the neutron star. 

We solve the diffusion equation for an extended disk with an initial  surface-density distribution in the power-law form, $\Sigma(r)\propto r^{-3/4}$, which is the characteristic surface-density profile for  steady thin accretion disks (see e.g. Frank, King \& Raine 2002). At a given time, the temperature in the disk decreases with increasing radial distance $r$ from the center. At a given $r$, the temperature decreases with time because of  fallback disk evolution and decreasing irradiation strength.In the calculations, for numerical reasons, we start with $\rout = 5 \times 10^{14}$ cm. After the first time step, the radius at which the disk temperature is as low as the minimum temperature $\Tp$ for the disk to be active is identified as the dynamical outer radius, $\rout$, of the viscously active disk. As the temperatures, starting from the outermost disk, decrease below  $\Tp$, $\rout$ propagates inward. The X-ray irradiation of the disk is important  in the evolution of the neutron star, since it prolongs the lifetime of the disk by delaying the inward propagation of $\rout$. X-ray irradiation flux can be written as $\Firr = C \Mdot c^2 / (4 \pi r^2$)  (Shakura \& Sunyaev 1973) where $c$ is the speed of light.  From the results of our earlier work on the IR and X-ray  emission of persistent AXP/SGRs (Ertan \& \Caliskan 2006), we found the value of the  irradiation efficiency  $C$ to be in the range of $\sim 1 - 7 \times 10^{-4}$  which is similar to the range of $C$ estimated for low-mass X-ray binaries  (Dubus et al. 1999).  There is a degeneracy between the parameters $C$ and $\Tp$. With two extreme values of  $C$,  similar evolutionary curves  can be obtained by changing $\Tp$ only by a factor of $\sim 1.6$. The range of $C$ obtained in our earlier work constrains  $\Tp$ to below $\sim 200$ K (Ertan et al. 2009), consistent with the results of Inutsuka \& Sano (2005). 

Accretion onto the surface of the neutron star is the dominant source of the X-ray luminosity. The accretion rate, $\Mdot$, is related to the X-ray luminosity through $\Lx = G M \Mdot / R$ where $R$ is the radius of the neutron star. In the fallback disk model of AXP/SGRs, the sources are in the propeller regime and a fraction of the matter arriving at the inner disk radius is accreted onto the star (Alpar 2001; Chatterjee, Hernquist \& Narayan 2000). This can be written as $\Mdot = \eta \Mdotin$ where $\eta \leq 1$.  For simplicity,  we take $\eta = 1$ in all our calculations. Similar results can be obtained with $\eta < 1$. Depending on the initial conditions, the sources can enter the accretion phase at different epochs of the evolution. Some sources may never enter the accretion phase and probably continue their evolution as radio pulsars until their rotational power is no longer sufficient to produce pulsed radio emission. Other sources, with different initial conditions, may evolve through an initial pulsar phase into a subsequent accretion epoch. When accretion is not possible, that is, when the inner disk cannot penetrate the light cylinder, the X-ray emission is mainly due to intrinsic cooling of the star. For the cooling luminosity, $\Lcool$, we use the theoretical cooling curves calculated for the neutron stars with conventional magnetic dipole fields (Page 2009). In the luminosity calculation, in addition to $\Lcool$ we also include the intrinsic dissipative heating of the neutron star under the dipole and disk torques acting on the star (Alpar 2007).   

We use the $\a$ prescription for the kinematic viscosity $\nu = \a c_s h$ where $c_s$ and $h$ are the local sound speed and the pressure scale-height of the disk respectively (Shakura \& Sunyaev 1973). In the long-term evolution of the disk, the mass-flow rate from the outer to the inner disk is determined by the viscosities in the cold outer disk. Following the results of the detailed work on the X-ray enhancement light curves of AXP/SGRs (\Caliskan \& Ertan 2012) we take $\a = 0.045$.  

The disk could remain stable (i.e., not blown away by radiation pressure) when $\rlc < \rin$ for a certain range of $\rin$ values depending on the angle $\theta$ between the magnetic dipole and the rotation axes of the neutron star (Ek\c{s}i \& Alpar 2005). The maximum $\rin$ for the disk to remain stable is a few $\rlc$ for large $\theta$, while the disk is stable for all $\rin$ values if the axes are aligned (Ek\c{s}i \& Alpar 2005). Even if the disk remains stable, it cannot apply an efficient torque on the star if the inner disk loses contact with the closed field lines. An alternative and stronger possibility in this regime is that the inner disk matter cannot be propelled efficiently from the system and piles up at the inner disk.  In this case, because of the increasing surface-density gradient at the inner disk, $\rin$ approaches to $\rlc$  and a steady state is reached with an inner disk that remains in contact with the closed dipole field lines. The system then follows a tracking phase with $\rin \sim \rlc$ (Ertan et al. 2009; Alpar et al. 2011; \Caliskan et al. 2013). As in our earlier work, we adopt that the inner disk follows this tracking phase when the inner disk cannot penetrate the light cylinder. 

In the accretion regime, the disk spin-down torque acting on the star can be written as $N = \dot{M} (G M \rA)^{1/2}~ F(\fast)$ where the dimensionless torque $F(\fast)$ is a function of the fastness parameter $\fast = \Omegastar/\OmegaK(\rA)$ where  $\OmegaK(\rA)$ is the angular frequency of the disk at $r = \rA$. From our earlier analysis, we found that the dimensionless torque in the form $F(\fast) = \beta (1-\fast^p)$ can produce the period evolution of AXP/SGRs with $\beta = 0.5$ and $p = 2$ (Ertan \& Erkut 2008; Ertan et al. 2009). Substituting $F(\fast)$ in the torque equation, we have
\be
 N =  \frac{1}{2} ~\Mdotin ~(G M \rin)^{1/2}~ (1-\fast^2) = I~ \dot{\Omegastar}
\ee
where $I$ is the moment of inertia of the neutron star. When $\rA < \rlc$, we calculate $N$ with $\rin = \rA$ using equation (1).  When $\rA$ is found to be greater than $\rlc$, we take $\rin = \rlc$.   

In the spin-down with accretion phase ($\rco < \rin < \rlc$), substituting equation (1) in equation (2), the period derivative, $\Pdot$, of the neutron star is found to be independent of both $\Mdot$ and $P$ when $\fast$ is sufficiently beyond unity. This is probably the case for most of the persistent AXPs and SGRs currently in the accretion phase except for a few sources which are likely to be close to rotational equilibrium. 

The \textquotedblleft low-B magnetars\textquotedblright~SGR 0418+5729 and Swift J1822.3$-$1606, which seem to have completed the accretion epoch, are evolving in the tracking phase with $\rin \simeq \rlc$ (Alpar et al. 2011; Benli et al. 2013). In this phase, the disk torque is proportional to $\Mdotin$. The ratio of the ram pressure to magnetic pressure at $\rlc$, and the disk torque decrease with time. By the time the magnetic dipole torque dominates the disk torque, the X-ray cooling luminosity of the source is already below the detection limits in most cases. In the tracking phase following the accretion phase, the pulsed radio emission rate depends on the dipole field strength and the period of the source (see Sections 3 and 4). We assume that magnetic field decay is negligible within the observable timescale of the AXP/SGRs and XDINs ($\tau \sim 10^6$ yr). Sources with appropriate initial conditions could start their evolution in the tracking phase. These are likely to be observed as radio pulsars until they cross the radio pulsar \textquotedblleft death line\textquotedblright.

\section{RESULTS}

Out of seven XDINs, six sources have measured period and period derivatives\footnote{When we were submitting this paper the detection of the period ($\simeq 3.39$ s) and a tentative (2 $\sigma$) period derivative $(\sim 1.6 \times 10^{-12}$ s s$^{-1}$) was reported for the seventh source RX J1605.3+3249 (Pires et al. 2014). We intend to investigate the evolutionary possibilities of this source when the period derivative is confirmed.}. For comparison with our model results, we have converted the reported X-ray fluxes of the six XDINs with known $P$ and $\Pdot$ into the unabsorbed fluxes and luminosities using the distances, $\NH$ values and blackbody temperatures given in the corresponding papers. There are large uncertainties in the distances of some XDINs which are reflected in the uncertainties of the X-ray luminosities. Corrections to luminosities could require modification of our model parameters reported here. 

For RX J0420.0-5022, $P = 3.453$ s (Haberl et al. 2004) and $\Pdot =2.8\pm0.3 \times 10^{-14}$ s s$^{-1}$ (Kaplan \& van Kerkwijk, 2011). The distance of this source was estimated as $\sim 345$ pc by Posselt et al. (2007). The $0.1-2.4$ keV observed flux was reported as $\sim 5 \times 10^{-13}$ erg s$^{-1}$ cm$^{-2}$ (Haberl et al. 2004). Using the properties of their spectral fit, we calculate the bolometric luminosity as $\Lx \simeq 2.6 \times 10^{31}$ erg s$^{-1}$.

The period of RX J1308.6+2127 (RBS 1223) is $P = 10.31$ s (Haberl 2004). The most recent timing analysis gives  $\Pdot = 1.120(3) \times 10^{-13}$ s s$^{-1}$ (Kaplan \& van Kerkwijk 2005). The distance is estimated to be $76 - 380$ pc by Schwope (2005) and 380 $^{+15}_{-30}$ pc by Hambaryan et al. (2011). Using $d = 380$ pc and the observed flux reported by Schwope, Schwarz \& Greiner (1999), we obtained $\Lx = 7.9 \times 10^{31}$ erg s$^{-1}$.

For RX J0806.4-4123, $P = 11.37$ s and $\Pdot = 5.5(30) \times 10^{-14}$ s s$^{-1}$ (Kaplan \& van Kerkwijk 2009a). The distance of this source was estimated as 235 - 250 pc by Posselt et al. (2007). Using $d = 240$ pc and with the observed X-ray flux reported by Haberl et al. (2004), we find $\Lx = 2.5 \times 10^{31}$ erg s$^{-1}$.

\begin{figure}
\includegraphics[height=.35\textheight,angle=270]{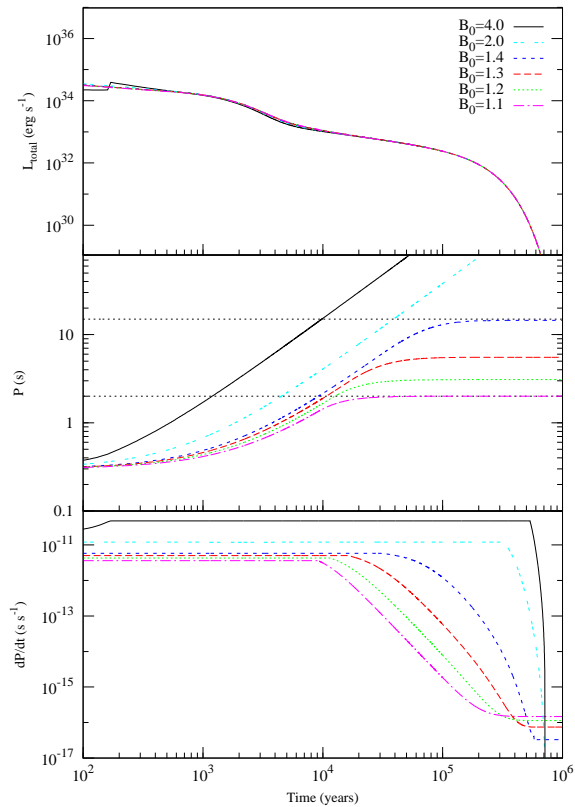}
\caption{Total luminosity, period and period derivative evolution of the model sources with different magnetic dipole fields. For all model sources $\Md = 3 \times 10^{-6} \Msun $. The magnitudes of the magnetic dipole field on the pole of the star, $B_0$, are given in units of $10^{12}$ G in the top panel (see the text for a discussion). } 
\end{figure}

The period of RX J1856.5-3754 is 7.055 s (Tiengo \& Mereghetti 2007) and $\Pdot = 2.97(7) \times 10^{-14}$ s s$^{-1}$ (van Kerkwijk \& Kaplan 2008). The distance is highly uncertain. Estimates range from 123$^{+11}_{-15}$ pc (Walter et al. 2010) to 167$^{+18}_{-15}$ pc (Kaplan, van Kerkwijk \& Anderson 2007). Posselt et al. (2007) estimate the distance as 135$ \pm $25 pc. With $d = 135$ pc, $\Lx$ is found between $\sim 5 \times 10^{31}$ erg s$^{-1}$ and $\sim 1 \times 10^{32}$ erg s$^{-1}$ (Walter \& Lattimer 2002; Pons et al. 2002; Drake et al. 2002; Burwitz et al. 2003). We take $\Lx$ = $9.5 \times 10^{31}$ erg s$^{-1}$ obtained by Burwitz et al. (2003) with the lowest $\chi^2$.

For RX J2143.0+0654 (RBS 1774), the timing analysis by Kaplan \& van Kerkwijk (2009b) gives $P$ = 9.428 s and $\Pdot = 4.1(18) \times 10^{-14}$ s s$^{-1}$. The upper limit to the distance  was given as 300 pc by Posselt, Neuh\"{a}user \& Haberl (2009) and as 390$-$430 pc by Posselt et al. (2007). The unabsorbed X-ray fluxes reported by Kaplan \& van Kerkwijk (2009b), Zane et al. (2005) and Rea et al. (2007) are 4.8, 6.1 and 5.6 $\times 10^{-12}$ erg s$^{-1}$ cm$^{-2}$ respectively. Using the flux reported by Rea et al. (2007), we find  $\Lx \simeq 1.1 \times 10^{32}$ erg s$^{-1}$ with d = 400 pc.

The period of RX J0720.4-3125 is $P$ = 8.39 s (Haberl et al. 1997). The most recent analyses give $\Pdot \sim 7 \times 10^{-14}$ s s$^{-1}$ (van Kerkwijk et al. 2007; Hohle et al. 2010). The distance was estimated as 235 $-$ 270 pc by Posselt et al. (2007) and as 360$^{+170}_{-90}$ by Kaplan, Kerkwijk \& Anderson (2007). We use d = 270 pc in our calculations, since it agrees with both estimates. The X-ray flux was reported as (9 $\pm~2) \times 10^{-12}$ erg s$^{-1}$ cm$^{-2}$ by Kaplan et al. (2003), which corresponds to a bolometric luminosity of $\sim 1.6 \times 10^{32}$ erg s$^{-1}$.

In the present work, we have investigated the long-term evolution of these six XDINs (Table 1).  The present day X-ray luminosities, periods and period derivatives of these sources can be produced simultaneously by the neutron star's evolution with a fallback disk. For a neutron star with a magnetic dipole field $\sim 10^{12}$ G to evolve into the observed period range of XDINs ($3 - 12$ s)  the system should pass through an accretion phase. Our results indicate that these sources slowed down to long periods during accretion epochs lasting from  a few $10^4$ yr to a few $10^5$ yr. They completed their accretion phases in the past, and now they are in the tracking phase, still spinning down by disk torques, which are weaker than in the accretion epoch. At present, the X-ray luminosities of these sources are likely to be powered by the intrinsic cooling of the neutron star. If the inner disk had never penetrated into the light cylinder the disk torque could not have spun down a neutron star with a conventional $\sim 10^{12}$ G dipole field to XDIN periods within the cooling timescale, $\tau_{\mathrm{cool}} \sim 10^6$ yr (e.g. Page 2009). 

\begin{figure}
\includegraphics[height=.35\textheight,angle=270]{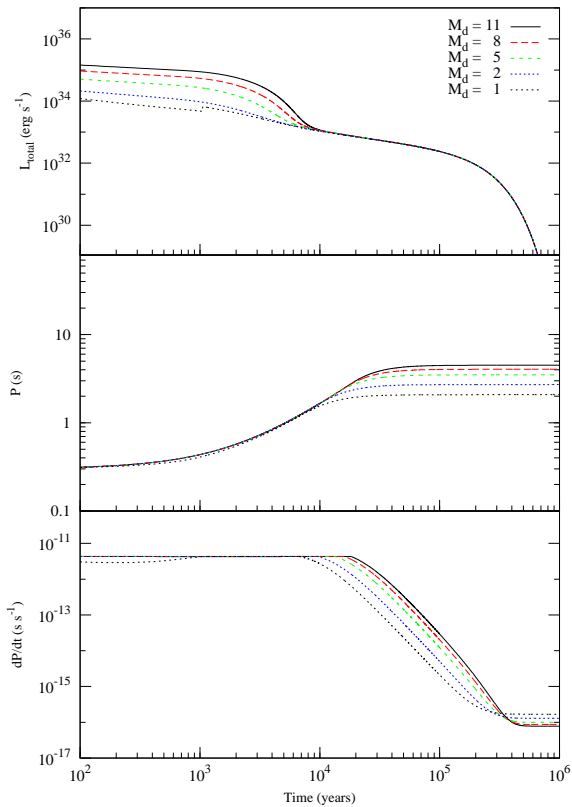}
\caption{The luminosity, period and period derivative evolution of the model sources with different initial disk masses. The magnetic dipole field is the same ($B_{0}$ = 1.2 $\times$ 10$^{12}$ G) for all sources. The disk masses are given in units of $10^{-6} \Msun$ in the top panel. } 
\end{figure}

We present illustrative model curves in Figures 1 and 2 to show the dependence of evolutionary paths on the disk mass, $\Md$, and the dipole field strength on the pole of the star, $B_0$, respectively. Different phases of evolution can be followed from the $\Pdot$ curves given in the bottom panels of Figures 1 and 2. In the accretion phase ($\rco < \rin <\rlc $), from Equation (1), it is found that $\Pdot$ is proportional to $B_0^2$, and independent of $\Mdotin$ and $P$. That is, the evolutionary phases with constant $\Pdot$ correspond to accretion epochs. In this phase, for a given $\Md$, the sources with higher $B_0$ reach longer periods. After the accretion phase, the system enters the tracking phase in which $\rin \simeq \rlc$. The disk torque decreases with decreasing $\Mdotin$ (the mass inflow rate arriving at $\rin \simeq \rlc$). Accretion onto the neutron star has stopped, $\Mdot = 0$, and the X-ray luminosity is supplied by the intrinsic cooling of the neutron star. In Figure 1, we also give illustrative model curves that could represent the evolution of AXP/SGRs (the two upper curves). These model sources with relatively high $B_0$ values cannot acquire the properties of XDINs. Our results indicate that XDINs could be distinguished from AXP/SGRs by having weaker $B_0$ fields.  
    
In Figure 2, we give the evolutionary curves for different disk masses keeping the initial dipole magnetic field constant at $B_0 \simeq 1.2 \times 10^{12}$ G. It is seen that systems with greater $\Md$ have longer accretion phases. We obtain these illustrative model curves with $\Tp \sim 200$ K and $C = 7\times 10^{-4}$. Using $C = 1\times 10^{-4}$, evolutionary curves similar to those in Figure 2 can be obtained with $\Tp \simeq 125$ K. Even with the highest reasonable value of $C$, the model curves obtained with $\Tp > 300$ K do not lead to XDIN or AXP/SGR properties for any disk mass. 

\begin{table*}[t]
\caption{\label{table:noktaatisi}The ages and the disk parameters for the six XDINs. The ages  are constrained by the theoretical cooling luminosity (Page 2009) and the estimated bolometric X-ray luminosities.}

\begin{minipage}{\linewidth}
\begin{center}
\begin{tabular}{c|c|c|c|c|c} \hline \hline
   & B$_{0}$ (10$^{12}$ G) & M$_{disk}$ (10$^{-6}$ M$_{\odot}$) & T$_{p}$ (K) & C (10$^{-4}$)& age (10$^{5}$ y)\\ \hline
RX J0720.4-3125 & 1.1 - 1.3  & 0.8 - 12  & 106 & 1 & 1.45 \\
RX J1856.5-3754 & 0.9 - 1.1   & 0.8 - 18  & 100 & 1 & 1.85 \\
RX J2143.0+0654 & 1.0 - 1.2  & 1.0 - 12  & 100 & 1 & 1.9 \\
RX J1308.6+2127 & 0.9 - 1.0  & 0.6 - 18  & 100 & 1.5 & 2.1 \\
RX J0806.4-4123 & 0.8 - 0.9  & 0.5 - 18  & 100 & 2.3 & 3.1 \\
RX J0420.0-5022 & 0.35 - 0.38  & 4.8 - 18  & 82 & 7 & 3.2 \\
\hline \hline
\end{tabular}\\

\end{center}
\end{minipage}
\end{table*}

Comparing Figures 1 and 2, it is seen that the dipole magnetic field strength $B_0$, rather than the disk mass, is the more effective initial condition for determining the long-term rotational evolution of the sources. In Figure 1, $\Md$ is the same for all model sources, and the $B_0$ values vary only by a factor of 4. We see that the model curves trace rotational properties of  AXP/SGR and XDINs from the shortest to the longest observed $P$ and $\Pdot$ values. For a given $B_0$, different $\Md$ values do not yield significantly different evolutionary paths. In Figure 2, disk masses change by an order of magnitude, while the final periods at $t = 1$ Myr remain in the 2 $-$ 5 s range. The $\Pdot$ curves of the sources with the same $B_0$  but different $\Md$ are also very similar. In the accretion phase, the accretion power is the dominant source of the X-ray luminosity, and depends on the initial disk mass $\Md$.   

With more detailed analysis, we also tried to reproduce the individual properties of XDINs with known $P$ and $\Pdot$ values. The evolutionary tracks of these sources are seen in Figure 3. Since all these XDINs are currently powered by the intrinsic cooling luminosity of their neutron stars, observed luminosities and the theoretical cooling curves constrain the ages of the model sources. In all these calculations, we first take $C = 1 \times 10^{-4}$ and $\Tp$ = 100 K. In some cases, we obtained better fits with slightly different $C$ and $\Tp$ values. The model curves given in Figure 3 are obtained with $\Tp$ values that remain in a narrow range between 80 and 110 K. Assuming that AXP/SGRs and XDINs have similar disk compositions, we expect that they have similar critical temperatures. The irradiation efficiency might change with the accretion rate; it is expected to be similar for sources in the same accretion regime. We perform simulations tracing all possible values of $\Md$ and $B_0$ with $P_0 = 300$ ms. For the model sources that can enter the accretion phase, the source properties at the end of this phase are not sensitive to $P_0$ (Ertan et al. 2009). For all XDINs, the $B_0$ values that can produce the reasonable evolutionary curves remain in the $\sim 0.3 - 1.3 \times 10^{12}$ G range. The ranges of model parameters that can produce the individual properties of the six XDINs are listed in Table 1. 

\begin{figure}
\includegraphics[height=.35\textheight,angle=270]{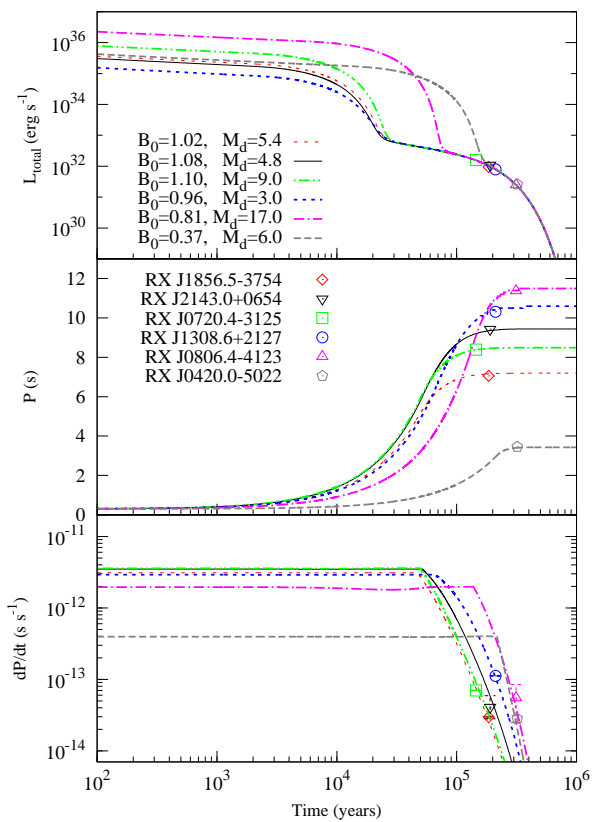}
\caption{The model curves that can simultaneously produce the luminosity, period and period derivative of the six XDINs. For each source, we determined the ranges of $\Md$ and $B_0$ that can produce the source properties (see Table 1). Here, we show illustrative model curves that can represent the long-term evolution of these XDINs. The values of $B_0$ and $\Md$ used in the models are given in the top panel in units of $10^{12}$ G and $10^{-6} \Msun$ respectively. }
\end{figure}

\begin{figure}
\includegraphics[height=0.24\textheight,angle=0]{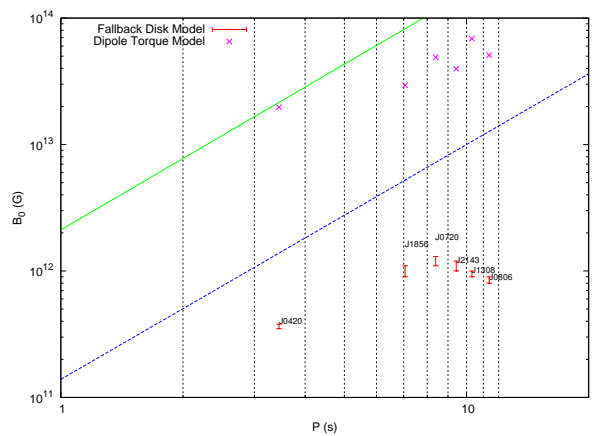}
\caption{Magnetic dipole field strength on the pole of the neutron star versus period distribution for six XDINs. The two parallel lines (green and blue) show the upper and lower bounds of the radio pulsar death valley from Chen \& Ruderman (1993). Crosses are the $B_0$ values inferred from the dipole torque formula. Vertical bars show the ranges of $B_0$ that can produce the properties of the sources in the fallback disk model (see Figure 3). It is seen that all these sources remain below the lower boundary of the death valley (death line) indicating that these XDINs cannot be normal radio pulsars, if they are evolving with fallback disks. }
\end{figure}

Our results imply that all six XDINs completed the long-term accretion phase. They are not accreting matter from the disk at present, while they are still being slowed down by the disk torques. Without accretion, these sources are free to emit pulsed radio emission. However, the dipole fields $B_0$ inferred in the fallback disk model are significantly weaker than those inferred from the dipole torque formula, so that at present the XDINs do not have sufficient voltages to sustain pulsed radio emission. Allowed ranges of $B_0$ values for each XDIN are given in Table 1 and plotted in Figure 4.  All these  sources are indeed well below the lower border of the radio pulsar death valley in the extended galactic population, and are not expected to show radio pulses. Younger XDINs with shorter periods should be able to emit radio pulses, but they would still be evolving in the long-term accretion phase during which mass-flow onto the neutron star hinders the radio emission. Once the accretion epoch is over, XDINs have weak rotational power and are not likely to produce radio emission.

The inner disk conditions of XDINs in the efficient propeller phase could be rather different from those of persistent AXPs in the accretion phase, for which the boundary layer extends down to the co-rotation radius in a steady state with magnetic heating in the boundary layer negligible compared to viscous dissipation (Rappaport, Fregeau \& Spruit 2004). By contrast, in the propeller phase, magnetic heating in the boundary layer could be a significant extra heating source in addition to viscous dissipation and irradiation. This additional heating in the propeller phase can supply the difference ($\sim 2-3$ orders of magnitude) between  the observed R band flux (Kaplan et al. 2011) and the model calculation, for an irradiated  face on thin disk. A calculation of the magnetic heating is beyond the scope of the present work; a detailed study is in progress and will be presented in a separate paper.

\section{DISCUSSION AND CONCLUSIONS}

We have examined the long-term evolution of the six XDIN sources, with measured $\Pdot$ out of the current total 7 XDINs. First, we find that the fallback disk model is successful in producing the observed individual properties of all these sources. XDIN properties are obtained after the source has gone through a past accretion phase with spin-down, with the fallback disk protruding into the neutron star's magnetosphere. At present, the fallback disk is tracking the light cylinder, with continuing spin-down under the disk torques, but no accretion onto the neutron star. The X-ray luminosities of XDINs are provided by cooling. Next, tracing the initial parameters, we have determined the allowed ranges of the dipole field strength on the pole of the star, $B_0$, and the disk mass, $\Md$, for each source (Table 1). Our results constrain the $B_0$ values of the six XDINs into the  $\sim 0.3 - 1.3 \times 10^{12} $ G range.  Comparing this range of dipole field strength with earlier results obtained by Ertan et al. (2009) for AXP/SGRs, we surmise that XDINs tend to have dipole fields weaker than those of most known AXP/SGRs. This result should be confirmed through further detailed work on the persistent and transient AXP/SGRs.  

Reasonable evolutionary model curves for XDINs can be obtained with a wide range of initial fallback disk masses, $\Md$, for all sources (Table 1). We could not test whether there is a correlation between $\Md$ and $B_0$, since our results do not constrain the disk masses. For the sources that do not accrete at present, like XDINs, it is not possible to constrain the initial mass of the disk in most cases. A possible $\Md - B_0$ correlation could be tested through further investigation of persistent AXP/SGRs which are powered by accretion onto the star. This analysis will help us understand the differences in the initial conditions of different young neutron star populations. 

All known XDINs are close-by objects within $\sim 400$ pc of the Sun. Statistical analysis  considering the properties of the Gould Belt gives a galactic birth rate of $\gtrsim$ 1 century$^{-1}$ (Popov et al. 2006; see also Section 1). This raises a critical question for the models: Only the XDINs in the solar neighborhood are observable in X-rays. The inferred large galactic population should be detectable in the radio band if they are active radio pulsars.  If these sources evolve in vacuum without fallback disks, their dipole field on the pole of the star can be estimated as $B_0 \simeq 6.4 \times 10^{19} \sqrt{P \Pdot}$ G. Observed $P$ and $\Pdot$ values give $B_0 \sim 10^{13} - 10^{14}$ G. It is expected to see these sources as radio pulsars, if XDINs indeed slow down by magnetic dipole torques with these strong fields. Five out of seven XDIN sources, located within 0.4 kpc, have periods greater than 7 s. Among the currently known radio pulsars in the ATNF pulsar catalog (Manchester et al. 2005), there is only a single radio pulsar with $P > 7$s ($P = 7.7$ s) and $B > 10^{13}$ G which is located at a distance of 3.3 kpc. We are not sure whether this can be  accounted for  with the selection effects and even the most conservative beaming fraction formulas for the long-period systems, if XDINs and radio pulsars with large dipole magnetic fields are taken to be members of the same population. A detailed population synthesis or statistical analysis will not be attempted here.

This problem encountered in the dipole torque model is naturally resolved in the fallback disk model. During the initial long-term accretion phase radio emission is suppressed due to mass-flow onto the star. Is it possible to observe pulsed radio emission from these sources after the mass accretion terminates?  The neutron stars that evolve like the six XDINs we studied here are not likely to produce beamed radio emission when accretion stops, because: (1) their dipole fields are much weaker than inferred from the dipole torque formula (Table 1), and (2) at the end of the accretion phase, they have already attained long periods. In Figure 3, accretion phases correspond to the epochs with constant $\Pdot$ given in the lower panel. The period at the end of the accretion phase corresponds to a location in the $B_0 - P$ plane which is already below the lower border of the death valley. Subsequent evolution moves the source even further away from the death valley, to the present locations shown in Figure 4.    

To sum up: (1) The rotational properties and the X-ray luminosities of XDINs can be explained by the fallback disk model that was employed earlier to explain the general properties of AXP/SGRs (Figure 1). The main disk parameters used in the present work for different XDIN sources are similar to each other and also to those used in the long-term evolution models of AXP/SGRs (Table 1). (2) The distinction between XDINs and AXP/SGRs seems to be that the former have weaker dipole magnetic fields. Finally,  (3) the model can also account for non-detection of many radio pulsars evolving to the properties of known XDINs. A neutron star with a conventional dipole field must have passed through the long-term accretion phase to acquire the periods of  known XDINs within the cooling timescale. In the accretion phase, the radio emission is quenched by the mass flow onto the neutron star.  After termination of the accretion phase, the sources no longer have sufficient rotational power for radio emission.

\section*{Acknowledgments}

We acknowledge research support from Sabanc\i\ University, and from
T\"{U}B{\.I}TAK (The Scientific and Technological Research Council of
Turkey) through grant 113F166. M.A.A. is a member of the Science Academy - Bilim Akademisi, Istanbul, Turkey.

\bsp

\label{lastpage}


\begin{thebibliography}{99}
\bibitem[\protect\citeauthoryear{}{2001}]{b1} Alpar M. A. 2001, ApJ, 554, 1245
\bibitem[\protect\citeauthoryear{}{2007}]{b2} Alpar M. A. 2007, Ap\&SS, 308, 133
\bibitem[\protect\citeauthoryear{}{2011}]{b3} Alpar M. A., Ertan \"U., \Caliskan \c{S}., 2011, ApJ, 732, L4
\bibitem[\protect\citeauthoryear{}{1991}]{b4} Balbus S. A., Hawley, J. F. 1991, ApJ, 376, 214
\bibitem[\protect\citeauthoryear{}{2013}]{b5} Benli O., \Caliskan \c{S}., Ertan \"U., Alpar M.~A., Tr{\"u}mper J.~E., Kylafis N.~D. 2013, ApJ, 778, 119
\bibitem[\protect\citeauthoryear{}{2003}]{b6} Burwitz V., Haberl F., Neuh\"{a}user R., Predehl P., Tr{\"u}mper J., Zavlin V.~E., 2003, A\&A, 399, 1109
\bibitem[\protect\citeauthoryear{}{2000}]{b7} Chatterjee P., Hernquist L., Narayan R., 2000, ApJ, 534, 373
\bibitem[\protect\citeauthoryear{}{1993}]{b8} Chen K., Ruderman M. 1993, ApJ, 402, 264 
\bibitem[\protect\citeauthoryear{}{2013}]{b9} \c{C}al{\i}\c{s}kan \c{S}., Ertan \"{U}., Alpar M. A., Tr\"umper J. E., Kylafis N. D., 2013, MNRAS, 431, 1136
\bibitem[\protect\citeauthoryear{}{2012}]{b10} \c{C}al{\i}\c{s}kan \c{S}., Ertan \"{U}., 2012, ApJ, 758, 98
\bibitem[\protect\citeauthoryear{}{1973}]{b11} Davidson K., Ostiker J. P., 1973, ApJ, 179, 585   
\bibitem[\protect\citeauthoryear{}{2002}]{b12} Drake J. J. et al. 2002, ApJ, 572, 996
\bibitem[\protect\citeauthoryear{}{1999}]{b13} Dubus, G., Lasota J. P., Hameury J. M., Charles P. 1999, MNRAS, 303, 139
\bibitem[\protect\citeauthoryear{}{2005}]{b14} Ek\c si K. Y., Alpar M. A., 2005, ApJ, 620, 390
\bibitem[\protect\citeauthoryear{}{2006}]{b15} Ertan \"{U}, \c{C}al{\i}\c{s}kan \c{S}. 2006, ApJ, 649, L87 
\bibitem[\protect\citeauthoryear{}{2008}]{b16} Ertan \"{U}., Erkut M. H. 2008, ApJ, 673, 1062
\bibitem[\protect\citeauthoryear{}{2007}]{b17} Ertan \"{U}., Erkut M. H., Ek\c si K. Y., Alpar M. A. 2007, ApJ, 657, 441
\bibitem[\protect\citeauthoryear{}{2009}]{b18} Ertan \"{U}., Ek\c si K. Y., Erkut M. H., Alpar M. A. 2009, ApJ, 702, 1309
\bibitem[\protect\citeauthoryear{}{2011}]{b19} Espinoza C.~M., Lyne A.~G., Kramer M., Manchester R.~N., Kaspi V.~M., 2011, ApJ, 741, L13
\bibitem[\protect\citeauthoryear{}{2002}]{b20} Frank J., King A., Raine D.~J., 2002, Accretion Power in Astrophysics, Cambridge University Press, pp.~398
\bibitem[\protect\citeauthoryear{}{2000}]{b21} Grenier I. A., 2000,  A\&A, 364,  L93 
\bibitem[\protect\citeauthoryear{}{2004}]{b22} Haberl F., 2004, Advances in Space Research, 33, 638 
\bibitem[\protect\citeauthoryear{}{2007}]{b23} Haberl F., 2007, Ap\&SS, 308, 181 
\bibitem[\protect\citeauthoryear{}{1997}]{b24} Haberl F., Motch C., Buckley D. A., Zickgraf F.-J., Pietsch W., 1997, A\&A, 326, 662
\bibitem[\protect\citeauthoryear{}{2004}]{b25} Haberl F. et al., 2004, A\&A, 424, 635
\bibitem[\protect\citeauthoryear{}{2011}]{b26} Hambaryan V., Suleimanov V., Schwope A. D., Neuh{\"a}user R., Werner K., Potekhin A.~Y., 2011, A\&A, 534, 74 
\bibitem[\protect\citeauthoryear{}{2010}]{b27} Hohle M. M., Haberl F., Vink J., Turolla R., Zane S., de Vries C.~P., M{\'e}ndez M., 2010, A\&A, 521, 11 
\bibitem[\protect\citeauthoryear{}{2005}]{b28} Inutsuka S., Sano T., 2005, ApJ, 628, L155 
\bibitem[\protect\citeauthoryear{}{2005}]{b29} Kaplan D. L., van Kerkwijk M. H. 2005, ApJ, 635, L65 
\bibitem[\protect\citeauthoryear{}{2009a}]{b30} Kaplan D. L., van Kerkwijk M. H. 2009a, ApJ, 692, L62
\bibitem[\protect\citeauthoryear{}{2009b}]{b31} Kaplan D. L., van Kerkwijk M. H. 2009b, ApJ, 705, 798
\bibitem[\protect\citeauthoryear{}{2011}]{b32} Kaplan D. L., van Kerkwijk M. H. 2011, ApJ, 740, L30
\bibitem[\protect\citeauthoryear{}{2003}]{b33} Kaplan D.L., van Kerkwijk M. H., Marshall H. L., Jacoby B.~A., Kulkarni S.~R., Frail D.~A., 2003, ApJ, 590, 1008 
\bibitem[\protect\citeauthoryear{}{2007}]{b34} Kaplan D. L., van Kerkwijk M. H., Anderson J. 2007, ApJ, 660, 1428
\bibitem[\protect\citeauthoryear{}{2011}]{b35} Kaplan D.~L., Kamble A., van Kerkwijk M.~H., Ho W.~C.~G. 2011, ApJ, 736, 117 
\bibitem[\protect\citeauthoryear{}{2010}]{b36} Kaspi V.~M. 2010, Proceedings of the National Academy of Science, 107, 7147
\bibitem[\protect\citeauthoryear{}{2009}]{b37} Keane E.~F., Kramer M. 2008, MNRAS, 391, 2009
\bibitem[\protect\citeauthoryear{}{2014}]{b38} Kylafis N.~D., Tr{\"u}mper J.~E., Ertan {\"U}., 2014, A\&A, 562, A62
\bibitem[\protect\citeauthoryear{}{1973}]{b39} Lamb F. K., Pethick C. I., Pines D. 1973, ApJ, 184, 271
\bibitem[\protect\citeauthoryear{}{2007}]{b40} Malofeev M. V., Malov O. I., Teplykh D. A., 2007, In Isolated Neutron Stars: From the Interior to the Surface, ed. S. Zane, R. Turolla, D. Page, Springer, Berlin Heidelberg, New York, 
\bibitem[\protect\citeauthoryear{}{2007}]{b41} Manchester R.~N., Hobbs G.~B., Teoh A., Hobbs M. 2005, VizieR Online Data Catalog, 7245, 0 
\bibitem[\protect\citeauthoryear{}{2008}]{b42} Mereghetti S., 2008, A\&ARv, 15, 225
\bibitem[\protect\citeauthoryear{}{2011a}]{b43} Mereghetti S., 2011a, High-Energy Emission from Pulsars and their Systems, Astrophysics and Space Science Proceedings, Springer-Verlag Berlin Heidelberg, p. 345
\bibitem[\protect\citeauthoryear{}{2011b}]{b44} Mereghetti S., 2011b,  Advances in Space Research, 47, 1317
\bibitem[\protect\citeauthoryear{}{2013}]{b45} Mignani R. P. et al., 2013, MNRAS, 429, 3517
\bibitem[\protect\citeauthoryear{}{2009}]{b46} Motch C., Pires A. M., Haberl F., Schwope A., Zavlin V. E., 2009, A\&A, 497, 423
\bibitem[\protect\citeauthoryear{}{2009}]{b47} Page, D., 2009, in "Neutron Stars and Pulsars", W. Becker, ed., Astrophysics and Space Science Library 357, 247
\bibitem[\protect\citeauthoryear{}{2014}]{b48} Pires A.~M., Haberl F., Zavlin V.~E., Motch C., Zane S., Hohle M. M., 2014, A\&A, 563, A50 
\bibitem[\protect\citeauthoryear{}{2002}]{b49} Pons J. A., Walter F. M., Lattimer M., Prakash M., Neuh{\"a}user R., An P., 2002, ApJ, 564, 981
\bibitem[\protect\citeauthoryear{}{2006}]{b50} Popov S. B., Turolla R., Possenti A., 2006, MNRAS, 369, L23
\bibitem[\protect\citeauthoryear{}{2009}]{b51} Posselt B., Neuh\"{a}user R., Haberl F., 2009, A\&A, 496, 533 
\bibitem[\protect\citeauthoryear{}{2007}]{b52} Posselt B., Popov S. B., Haberl F., Tr{\"u}mper J., Turolla R., Neuh{\"a}user R., 2007, Ap\&SS, 308, 171
\bibitem[\protect\citeauthoryear{}{2004}]{b53} Rappaport S.~A., Fregeau J.~M., Spruit H. 2004, ApJ, 606, 436 
\bibitem[\protect\citeauthoryear{}{2007}]{b54} Rea N. et al., 2007, MNRAS, 379, 1484
\bibitem[\protect\citeauthoryear{}{2012}]{b55}Rea N. et al. 2012, ApJ, 754, 27 
\bibitem[\protect\citeauthoryear{}{2013}]{b56} Rea N. et al., 2013, ApJ, 775, L34
\bibitem[\protect\citeauthoryear{}{2012}]{b57} Scholz P., Ng C.-Y., Livingstone M. A., Kaspi V.~M., Cumming A., Archibald R.~F, 2012, ApJ, 761, 66
\bibitem[\protect\citeauthoryear{}{1999}]{b58} Schwope A. D., Schwarz R., Greiner J., 1999, A\&A, 348, 861
\bibitem[\protect\citeauthoryear{}{2005}]{b59} Schwope A. D., Hambaryan V., Haberl F., Motch C., 2005, A\&A, 441, 597
\bibitem[\protect\citeauthoryear{}{1973}]{b60} Shakura N. I., Sunyaev R. A., 1973, A\&A, 24, 337
\bibitem[\protect\citeauthoryear{}{2010}]{b61} Tetzlaff N., Neuh\"auser R., Hohle M. M., Maciejewski G., 2010, MNRAS, 402, 2369 
\bibitem[\protect\citeauthoryear{}{2007}]{b62} Tiengo A., Mereghetti S. 2007, ApJ, 657, L101
\bibitem[\protect\citeauthoryear{}{2010}]{b63} Tr\"umper J. E., Zezas A, Ertan \"U.,  Kylafis N. D., 2010, A\&A, 518, A46
\bibitem[\protect\citeauthoryear{}{2013}]{b64} Tr\"umper J. E., Dennerl K., Kylafis N. D., Ertan \"U., Zezas A. 2013, ApJ, 764, 49
\bibitem[\protect\citeauthoryear{}{2009}]{b65} Turolla R., 2009, in "Neutron Stars and Pulsars", W. Becker, ed., Astrophysics and Space Science Library 357, 141
\bibitem[\protect\citeauthoryear{}{2011}]{b66}Turolla R., Zane S., Pons J.~A., Esposito P., Rea N. 2011, ApJ, 740, 105 
\bibitem[\protect\citeauthoryear{}{2008}]{b67} van Kerkwijk M. H., Kaplan D. L., 2008, ApJ, 673, L163 
\bibitem[\protect\citeauthoryear{}{2007}]{b68} van Kerkwijk M. H., Kaplan D. L., Pavlov G. G., Mori K. 2007, ApJ, 659, L149
\bibitem[\protect\citeauthoryear{}{2012}]{b69} Vigan{\`o} D., Pons J.~A. 2012, MNRAS, 425, 2487 
\bibitem[\protect\citeauthoryear{}{2013}]{b70} Vigan{\`o} D., Rea N., Pons J.~A., Perna R., Aguilera D.~N., Miralles J.~A. 2013, MNRAS, 434, 123 
\bibitem[\protect\citeauthoryear{}{2002}]{b71} Walter F. M., Lattimer J. M. 2002, ApJ, 576, L145
\bibitem[\protect\citeauthoryear{}{2010}]{b72} Walter F. M., Eisenbeis T., Lattimer J. M., Kim B., Hambaryan V., Neuh{\"a}user R., 2010, ApJ, 724, 669
\bibitem[\protect\citeauthoryear{}{2005}]{b73} Zane S., Cropper M., Turolla R., Zampieri L., Chieregato M., Drake J.~J., Treves A., 2005, ApJ, 627, 397 


\end{thebibliography}
\end{document}